\documentclass[onecolumn,noshowpacs,nofootinbib,colorlinks,hyperindex,unicode,11pt]{revtex4}
\usepackage{graphicx}
\usepackage{tikz}

\usepackage{amsfonts,amsmath,amssymb,tikz,accents}
\usepackage[eulergreek]{sansmath}
\usepackage{indentfirst,bigints}
\usepackage{upgreek}
\usepackage{bm,graphics,color}
\usepackage{feynmp-auto}
\usepackage{slashed}
\counterwithout{equation}{section}
\usepackage{pgfplots}
\pgfplotsset{compat=1.16}

\usepackage{textgreek}
\usepackage{color,xcolor}
\def\0{\mbox{\boldmath$\displaystyle\mathbb{O}$}}

\def\p{\partial}

\usepackage{float,xcolor,upgreek}
\usepackage{color} 
\newcommand{\beq}{\begin{eqnarray}}
\newcommand{\eeq}{\end{eqnarray}}
\newcommand{\bea}{\begin{eqnarray}}
\newcommand{\eea}{\end{eqnarray}}

\begin{document}

\title{{Duality as a Method to Derive a Gauge Invariant Massive Electrodynamics and New Interactions}}

\author{G. B. de Gracia}
\affiliation{Federal University of ABC, Center of Mathematics,  Santo Andr\'e, 09210-580, Brazil.}
\email{g.gracia@ufabc.edu.br}
\author{B.M.Pimentel}
\affiliation{Institute of Theoretical Physics, Sao Paulo State University, 01156-970, S\~ao Paulo,  Brazil }
\email{bruto.max@unesp.br}

%%%%%% Abstract %%%%%%

\begin{abstract}
Taking into account the recent developments associated with duality in physics, this article is focused on investigating the properties of a tensor generalization of the electrodynamics dual to the standard vector model even considering the full radiative corrections. A discussion on duality is extended for non-linearly interacting models. The alternative tensor structure allows a new local self-interaction and also a partial gauge invariant mass term. In this way, the renormalization conditions, a new interaction, Ward-Takahashi identities for all the sets of gauge symmetries, and the Schwinger-Dyson quantum equations are carefully provided to consistently characterize the propagating modes of the quantum system. The possibility of this gauge invariant massive extension is considered from the perspective of a generalized Stueckelberg procedure for higher derivative systems.
\end{abstract}

\maketitle

\section{Introduction}

\indent The issue of duality has important implications in physics. It is interesting to mention some well-established examples such as the particle-vortex web of dualities in $2+1$ dimensions \cite{xxxpv1,xxxpv2} with applications in the context of topological insulators \cite{nastase}, the case of three-dimensional QCD \cite{xxxqcd1,qcd}, the boson-fermion duality \cite{xxxboson,boson} and the well-established weak dualities in condensed matter \cite{xxxcond1,xxxcond2}.  \\
\indent In the case of planar quantum electrodynamics, there is the possibility of adding a topological charge that classifies different vortex solutions \cite{xxxtop1}. But the point is, which is the associated symmetry? In this dimension, there is a dual map relating the photonic model to a scalar field Lagrangian. Then, the freedom to perform constant shifts in this field is related to this topological charge.\\
\indent An important technique in the context of duality for field theories is the so-called master action approach, see \cite{xxxdual1,xxxdual2,xxxdual3}. One can also mention recent applications of this methodology in topological insulators  \cite{nastase} and also in the framework of dual theories for the graviton \cite{dual4}, for example. An important prototype case, with a high conceptual correlation to this paper, is the relation between the so-called self-dual model and the Maxwell-Chern-Simons (MCS) one. It is well known that QED in lower dimensions has severe infrared (IR) singularities\cite{frenkel}. Then, in the eighties, a topological mass term was considered providing an infrared cut-off for the theory in 2 + 1 space-time dimensions. Interestingly, for manifolds without boundaries, the gauge symmetry is preserved even in the presence of this new term   \cite{xxxsd1}.\\
\indent The resulting theory is the so-called Maxwell-Chern-Simons (MCS)
model, describing a massive excitation. After that, a self-dual spin 1 model was conceived by Townsend and collaborators \cite{xxxsd3}. Even
though this new theory describes a massive vector field without imposing gauge symmetry, it was realized that this model is related to the above topological massive model through a dual correspondence derived by the previously mentioned master action procedure \cite{xxxdeser}.\\
\indent Regarding the power of duality, there is also the fertile research field of anti-de Sitter/conformal field theory (AdS/CFT) correspondence \cite{xxxads1}, and also the correlated AdS/CMT approach in condensed matter physics \cite{xxxads2,ads3}.\\
\indent It is worth mentioning some analogous dual maps for non-linearly interacting systems, like \cite{xxxnl}, the reference \cite{maluf} for the interacting BF theory, the action for partially massless gravity theory \cite{bou}, and the Curtright-Freund model \cite{curt2}. The alternative dual tensor structure can also provide physical extensions due to the possibility of new couplings, see \cite{rham1} for Proca-like and \cite{heis} for new $2$-form interactions.  Regarding antisymmetric tensor fields, dual formulations of spin $0$ or spin $1$ bosonic theories can be described by the massless and the massive phases of the Kalb-Ramond model \cite{xxxkalb}, respectively. One can also cite the effective theory for low energy excitations in QCD \cite{xxxkam1,xxxkam2} and the model for a perfect fluid \cite{fluido}. These systems present discontinuities in the spin content at the massless limit as an implication of their reducible Hamiltonian constraint structures \cite{xxxpha, xxxreduc, hell}. Interestingly, the model studied here, based on a symmetric field, does not present this feature as proved by the Stueckelberg procedure \cite{xxxstu}. \\  
\indent In the present article, the first objective is to develop dual electrodynamics in terms of a higher derivative bosonic sector described by a symmetric rank $2$ tensor field whose non-interacting limit was first obtained in \cite{xxxd1}. The massless phase presents a Weyl symmetry capable of being associated with the standard $U(1)$ symmetry of the QED$_4$. It also presents an extra reducible local freedom due to the use of the symmetric tensor field representation. This higher-order theory was obtained by a triple master action presenting well-defined Hamiltonian properties, being also in compliance with unitarity \cite{xxxd2}.  We prove that the model reproduces all QED$_4$ radiative corrections if a well-defined prescription for external sources and the non-linear couplings, suggested by the Gaussian dual map, are considered. The renormalization process reveals that no counter-term with a non-standard form is necessary, implying a renormalizable theory in one-loop. As a counter-example, we highlight the specific problems faced by an analogous dual interacting spin $1$ theory of second derivative order. Regarding this model, although the fermionic self-energy reproduces the QED$_4$ well-known form even considering the full radiative corrections, the renormalization of the bosonic sector demands a new class of counter terms. Based on the latter, an interesting prototype procedure for non-renormalizable models can be derived. Moreover, we explicitly show that the new negative norm pole arising due to the radiatively generated fourth derivative terms can be consistently eliminated from the asymptotic states by a well-defined renormalization process \cite{merline}. This achievement can also contribute to improvements in renormalizability and the establishment of asymptotic completeness in quantum gravity. The negative norm excitations introduced by the higher derivative loop corrections and counter terms, associated with products of the curvatures, are avoided by an analogous mechanism \cite{merlin2,merlin3}.\\ 
\indent The second objective is the main one. After establishing the non-linear duality with vector QED$_4$, we look for new couplings and derive the associated degree of divergence. As we are going to show, after a judicious gauge fixation and an additional restriction on the measure, a well-defined self-interaction preserving the Weyl symmetry can be suitably derived. It leads to a completely local Lagrangian structure. Although not possible in the vector representation, we show that the higher-rank field structure guarantees these features. It also furnishes a hint for a new kind of mass term compatible with this mentioned Weyl symmetry, the one related to the covariant derivative prescription. This process violates the scalar part of the reducible symmetry. However, we prove that the bosonic sector keeps the spin $1$ degrees of freedom if a given constraint on the couplings is considered. Basically, this mass term can be obtained by a specific field redefinition that resembles the one introduced in the so-called WTDIFF gravity model leading to a wider set of local symmetries \cite{tdiff1,tdiff2,tdiff3}. \\ 
\indent Accordingly, the quantum equations of motion in the presence of a spinor field interaction are analyzed considering the information from a set of Schwinger-Dyson equations characterizing the complete radiative structure of the model, see some recent applications in the context of QCD \cite{sch1,sch2}. These results complement the previous conclusions by considering the full quantum content. In order to define the propagating modes, the so-called helicity decomposition is also performed. Moreover, the quantum Stueckelberg procedure is also provided. Due to the tensor structure, in order to define a mass term, one can choose to break the hidden scalar symmetry instead of the Weyl one, related to the U($1$) local freedom of the fermionic part. The well-defined behavior of this gauge invariant massive extension is closely dependent on a restriction on the spinor interaction. Besides the quantum equations of motion, it is also corroborated by the
analysis of the saturated amplitude as well as some other correlated unitarity constraints. The Ward-Takahashi identities for the whole set of symmetries are carefully obtained.  With this knowledge, one can further constrain the free parameters in such a way that the renormalization conditions can be also successfully imposed for this massive phase. Finally, it is also important to mention that the present analysis can be understood as a laboratory for future extensions associated with spin $2$ theories, more specifically, gravity models. One possibility is to apply the mechanism outlined here to use higher rank fields to derive new spin $2$ couplings or even a massive gravity invariant under full linearized diffeomorphisms, in accordance with the relativity principle.\\
\indent The paper is organized as follows. In Sec.\ref{sec2}, the triple master action from which our target model is derived is revisited. The Sec.\ref{sec3} is devoted to proving that a dual QED$_4$ can be consistently derived by using the bosonic tensor field. Then, Sec.\ref{sec4} is focused on using this tensor structure to look for new couplings and mass terms. The Sec.\ref{sec5} demonstrates that the chosen mass term leads to a violation of the spin $1$ nature for the case of arbitrary spinor interaction. However, this problem can be circumvented by a proper restriction of the external source structure\footnote{Furnishing also a criterium to establish the non-linear field dependent interactions. }. In order to count the local degrees of freedom, the helicity decomposition is performed in Sec.\ref{sec7} to define the true propagating modes. The natural next step is the analysis of constraints from the quantum equations of motion in Sec.\ref{sec8} in which a class of spinor sources is considered. In order to provide a renormalization analysis, the quantum Stueckelberg procedure for higher derivative theories is introduced in Sec.\ref{sec9}. It allows the study of Ward-Takahashi identities in Sec.\ref{sec10} associated with all the local symmetries, even the one broken by the mass term. Then, in Sec.\ref{sec11}, the renormalization of the bosonic two-point structure is provided. Finally, we conclude and present new perspectives in Sec.\ref{sec12}. The metric signature $(+,-,-,-)$ is used throughout.

.

\section{  Revisiting the Dual Map for Tensor and Vector spin $1$ models     }\label{sec2}

\indent  There is a master action that relates three different models describing a spin-$1$ particle \cite{xxxd1,xxxd2}. Its explicit form is displayed below

\begin{align}     S\big[A,H,\mathcal{N},T,J\big]=\int d^4x\Big[&-\sqrt{2}m\ H^{\mu \nu}\partial_\mu A_\nu+\frac{m^2}{2}A_\mu A^\nu+\frac{m^2}{2}\Big(H_{\mu \nu}H^{\mu \nu}-\frac{H^2}{3}          \Big)\nonumber \\&+\p^\mu \big[\mathcal{N}_{\mu \nu}+H_{\mu \nu}\big]^2+A_\mu J^\mu+H_{\mu \nu}T^{\mu \nu}                \Big]  \end{align}
 \indent Then, after completing the square in two of these fields, the action for the remaining one is obtained. Accordingly, a higher derivative tensor representation of a spin $1$ particle is achieved

\begin{align}     S\big[\mathcal{N},T,J\big]=&\int d^4x\Big[\frac{1}{2}\p^\alpha \mathcal{N}_{\mu \alpha}\big(\eta^{\mu \nu}(\Box+m^2)-\p^\mu \p^\nu       \big)\p^\alpha \mathcal{N}_{\nu \alpha}\nonumber \\& -\frac{\sqrt{2}}{m}T^{\mu \nu}\big( \eta_{\mu \nu}\p^\mu \p^\nu \mathcal{N}_{\mu \nu}-\p^\alpha\p_{(\mu}\mathcal{N}_{\nu)\alpha } \big)+\frac{T^2-T_{\mu \nu}^2}{2m^2}+\p^\mu \mathcal{N}_{\mu \nu}J^\nu\Big]\end{align}
\indent Considering an analogous procedure, the Proca action is recovered

\begin{align}     S\big[A,T,J\big]=\int d^4x\Big[ -\frac{1}{4}F_{\mu \nu}F^{\mu \nu}+\frac{m^2}{2}A_\mu A^\mu+A_\mu J^\mu-\frac{\sqrt{2}}{m}T^{\mu \nu}\big(\eta_{\mu \nu}\p_\mu A^\mu-\p_{(\mu}A_{\nu)}   \big)+\frac{T^2-T_{\mu \nu}^2}{2m^2}           \Big]\end{align}

\indent A spin $1$ tensor field  representation with second derivative order can be also derived by completing the square in the other fields

\begin{align}     S\big[H,T,J\big]=\int d^4x\Big[-\big(\p^\mu H_{\mu \nu}\big)^2    +\frac{m^2}{2}\Big(H_{\mu \nu}H^{\mu \nu}-\frac{H^2}{3}          \Big) +\frac{ J_\mu J^\mu}{2m^2}+\frac{\sqrt{2}}{m}\p^\mu H_{\mu \nu}J^\nu+H_{\mu \nu}T^{\mu \nu}     \Big]\label{model}\end{align}

\indent  By varying these actions with relation to the c-number sources $T_{\mu \nu}$ and $J_\mu(x)$, one can obtain a set of dual maps relating different field propagators. There are contact terms due to the presence of squared source terms that avoid a complete association between $H_{\mu \nu}(x)$ and the Proca field. \\
\indent On the other hand, there is a strong \footnote{Meaning that there are no contact terms.} dual map relating the vector field to $\mathcal{N}_{\mu \nu}(x)$. Our goal here is to prove that the renormalization difficulties associated with the weak dual field $H_{\mu \nu}$ can be overcome by the higher derivative theory for the massive case as well as for its massless limit \footnote{Associated to a dual relation with Maxwell theory.}. Also, it is possible to carefully introduce a gauge invariant mass term for the latter. For the massive phase, the mentioned map explicitly reads

\bea \mathcal{N}_{\mu \nu}(x)\longleftrightarrow \Big[ \eta_{\mu \nu}\p_\beta A^\beta(x)-\p_{(\mu}A_{\nu)}(x)  \Big] \quad ,\quad    A_\mu(x) \longleftrightarrow \p^\nu \mathcal{N}_{\nu \mu}(x)           \eea

\indent According to \cite{xxxd1}, the higher derivative massive field has a positive saturated amplitude, satisfying a necessary condition for unitarity. Its massless limit is also compatible with this constraint. Some contents derived from this dual map are extrapolated to the massless case, furnishing useful suggestions. In the next sections, we prove that they are indeed pertinent.\\
\indent The massless phase has two sets of local symmetries. The first one is associated with action invariance under the transformations below
\bea  \mathcal{N}^{\mu \nu}(x) \to \mathcal{N}^{\mu \nu}(x)+ \varepsilon^{(\mu \alpha \beta \gamma}\varepsilon^{\nu) \sigma \omega \varepsilon}\p_\alpha \ \p_\sigma \Lambda_{[\omega \varepsilon  ] [\beta \gamma  ]}(x)   \eea
with $\Lambda_{[\omega \varepsilon  ] [\beta \gamma  ]}(x)$ being an arbitrary field which is antisymmetric under the exchange of the indices inside the square brackets. This is a reducible symmetry transformation with its own zero modes.\\ 
\indent There is also invariance under Weyl transformations. This symmetry structure can be applied to derive a minimal coupling prescription for our theory when, besides the c-number sources, an interaction with fermionic fields is included. It plays the role of the $U(1)$ invariance in vector QED$_4$. Its associated transformation explicitly reads

\bea \mathcal{N}_{\nu \beta}(x) \to \mathcal{N}_{\nu \beta}(x)+ \eta_{\nu \beta}\ \Omega(x) \eea
With $\Omega(x)$ being an arbitrary scalar field.\\
\indent  A suitable gauge condition for the reducible symmetry is the following
\bea \Box^2 \mathcal{N}^{\nu \alpha}(x)+\p^\nu \p^\alpha \big(\p_\mu \p_\sigma \mathcal{N}^{\mu \sigma}(x)  \big)-2\Box \p_\mu \p^{(\nu}\mathcal{N}^{ \alpha) \mu}(x)=0\eea
\indent In order to fix the Weyl symmetry, the following condition is chosen
\bea \p_\mu \p_\nu \mathcal{N}^{\mu \nu}(x)=0\eea
in order to establish a relation with the Lorenz condition of QED$_4$. In accordance with the Dirac criteria \cite{dirac}, it is the most general condition allowing the obtainment of a Klein-Gordon equation for each component of the field $\p_\nu \mathcal{N}^{\mu \nu}(x)$. Moreover, the associated gauge fixing Lagrangian acquires the same derivative order as the kinetic terms.\\
\indent The gauge fixed action written in terms of the spin projectors has the form \footnote{ See appendix A.}

\bea S_M=\frac{1}{2}\int d^4x\mathcal{N}^{\mu \nu}\Big(-\frac{1}{2}\Box^2 P^{(1)}_{ss} +\tilde \lambda \Box^2P^{(0)}_{\omega \omega}+\lambda \big(\Box^4P^{(2)}_{ss}+\Box^4P^{(0)}_{ss}\big)   \Big)_{\mu \nu \alpha \beta}\mathcal{N}^{\alpha \beta} \nonumber \\         \eea

\indent Considering the spin projectors algebra and their completeness relation, the momentum space propagator can be derived

\bea \mathcal{G}_{\mu \nu \alpha \beta}(k)=\Big(\frac{1}{\lambda k^8} P^{(2)}_{ss}-\frac{2}{k^4}P^{(1)}_{ss}+ \frac{1}{\tilde \lambda k^4}P^{(0)}_{\omega \omega} + \frac{1}{ \lambda k^8}P^{(0)}_{ss}      \Big)_{\mu \nu \alpha \beta}         \eea

\indent Although the previous analysis was made in terms of Gaussian actions, these observations and suggestions can be useful even for the interacting case. In order to explicitly verify unitarity constraints in the case of spinor electrodynamics, we consider the following prescription for the external gauge boson states 
\bea \upepsilon_{\mu \nu}^r(k)=\frac{k_\mu \upepsilon_\nu^r(k)+k_\nu \upepsilon_\mu^r(k)       }{k^2} \eea
in accordance with the fact that  $\epsilon_{\mu \nu}^r(k) $ has no direct physical interpretation, since the propagating pole is associated with $k^\mu \epsilon_{\mu \nu}^r(k) $. The polarization vectors given above are transverse and obey the completeness relation  $\sum_{r}\epsilon_\mu^r(k)\epsilon_\nu^{*r}(k)=-\big(g_{\mu \nu}-\frac{k_\mu k_\nu}{k^2} \big)  $.  For the case of spinor currents,  every external bosonic line is contracted with a term $k_{(\mu}\gamma_{\nu)}$, in which $k_\mu$ denotes the momentum of the external boson, present in the structures associated with the $1$(PI) functions. Then, considering our prescription, the overall factor $\gamma_\nu \upepsilon^\nu_r(k)$ arises, being suitable to our duality prescription. Here, $\gamma_\mu$ denotes the so-called Dirac gamma matrices.

\section{Feynman Rules, Radiative Corrections and Unitarity}\label{sec3}

\indent The action for the dual tensor QED$_4$ with fermionic spinor interaction reads\footnote{In order to establish the spinor interaction, we are considering the suggestions from the constraints assumed for the external c-number sources. The latter are the ones used in the obtainment of Green functions from the path integral. According to the previous sections, the constraints are associated with the local symmetries.  }
\bea S_{eff.}=\int d^4x\Big(+i\bar \Psi\gamma^\mu \p_\mu \Psi-m_e\bar \Psi \Psi-e\p^\nu \mathcal{N}_{\nu \mu} \bar \Psi\gamma^\mu\Psi  +\frac{1}{2}\p^\alpha \mathcal{N}_{\mu \alpha}\big(\eta^{\mu \nu}\Box-\p^\mu \p^\nu       \big)\p^\alpha \mathcal{N}_{\nu \alpha}\nonumber \\+\frac{\tilde \lambda}{2}(\p^\nu \p^\alpha \mathcal{N}_{\nu \alpha})^2+\frac{ \lambda}{2}(\Box^2 \mathcal{N}^{\nu \alpha}+\p^\nu \p^\alpha \big(\p_\mu \p_\sigma \mathcal{N}^{\mu \sigma}  \big)-2\Box \p_\mu \p^{(\nu}\mathcal{N}^{ \alpha) \mu})^2 \Big)   \nonumber \\ \label{citar}\eea
in which $\Psi(x)$ denote a fermionic field.\\
\indent The action is invariant under the following local transformations 
\begin{align} \Psi(x)\to \Psi(x)+i\alpha(x)\Psi(x)\quad &,\quad  \bar \Psi(x)\to \bar \Psi(x)-i\alpha(x)\bar \Psi(x) \nonumber \\
\mathcal{N}_{\mu \nu }(x)\to \mathcal{N}_{\mu \nu }(x)+\frac{1}{e}\eta_{\mu \nu }\alpha(x)\quad& ,\quad N^{\mu \nu}(x) \to N^{\mu \nu}(x)+ \varepsilon^{(\mu \alpha \beta \gamma}\varepsilon^{\nu) \sigma \omega \varepsilon}\p_\alpha \ \p_\sigma \Lambda_{[\omega \varepsilon  ] [\beta \gamma  ]}(x)\end{align}
\indent The symmetry transformation associated with $\alpha(x)$ recovers the $U(1)$ structure of QED$_4$ since the action is written in terms of the divergence of the tensor field. This fact is the origin of the reducible symmetry under transformations parameterized by $\Lambda_{[\omega \varepsilon  ] [\beta \gamma  ]}(x)$  which has no analogous in the usual vector theory of QED$_4$.\\
\indent The Feynman propagator for the fermions is defined below \footnote{We are considering the standard notation $\slashed{A}=A_\mu\gamma^\mu $.}

        \bea \vcenter{\hbox{\includegraphics[width=3.2cm,height=1.15cm]{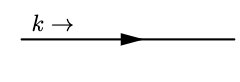}}}=\frac{i(\slashed{k}+m_e)}{\Big[ k^2-m^2_e+i\epsilon \Big]}\eea

and the previously defined bosonic propagator has the following definition
\begin{equation}
\vcenter{\hbox{\includegraphics[width=2.6cm,height=1.4cm]{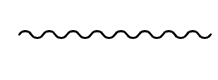}}}=i\mathcal{G}_{\mu \nu \alpha \beta}(k)
\end{equation}

\indent The interaction vertex has a derivative structure
\begin{equation}\vcenter{\hbox{\includegraphics[width=3.1cm,height=2.6cm]{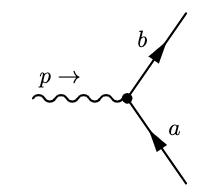}}}=ep_{(\mu}\gamma_{\nu)^{ab}}
       \end{equation}
       
       in which the orientation of the gauge boson momentum must be also taken into account.\\
\indent Since the diagrams have the same topology as in the vector QED$_4$ case, the relations \footnote{V denotes the number of vertices, $P_\gamma $ is the number of bosonic propagators, $P_e$ the number of fermionic propagators, with $N_\gamma$ and $N_e$ being the number of bosonic and fermionic external lines, respectively. $L$ denotes the number of loops.} $V=2P_\gamma +N_\gamma=P_e+\frac{1}{2}N_e$ and $L=P_e+P_\gamma-V+1$ are kept. The expression for the superficial degree of divergence becomes $D=4L+\big(V-N_\gamma \big)-P_e-4P_\gamma$, the coefficient $4$ for the bosonic internal line is due to the higher order structure of the propagator. The term  $\big(V-N_\gamma \big) $ expresses the fact that since the momentum is attached to the photon line, when it is internal, the vertex leads to an increase in the divergence order of the graph, but in the case in which it is external, there is no extra contribution to the divergence degree at all. Therefore, these constraints lead to
\bea D=4-N_\gamma-\frac{3}{2}N_e\eea
which is the same expression for the conventional formulation of QED$_4$.  \\
\indent The expression for the bosonic polarization tensor is the following

 \bea 
   i\mathcal{\pi}_{\mu \nu \alpha \beta}(p^2)=-e^2 tr \int \frac{d^4k}{(2\pi)^4}\frac{p_{(\mu}\gamma_{\nu)}(\gamma^\sigma(k_\sigma-p_\sigma)  +m_e)p_{(\alpha}\gamma_{\beta)}(\gamma^\rho k_\rho+m_e)}{\big( k^2-m^2_e\big)\big( (p-k)^2-m^2_e\big)}    \eea

see appendix B to access the definition and properties of the Dirac gamma matrices.\\
\indent The fulfillment of the following Ward-Takahashi-like identity for the polarization tensor is a heritage from the transverse nature of the usual QED$_4$ radiative correction associated with this specific derivative coupling structure \footnote{The massless limit of the Ward-Takashashi identities of \ref{sec10} recover the present system and indeed imply in this condition. }
\begin{align}\eta^{\mu \nu}\mathcal{\pi}_{\mu \nu \alpha \beta}(p^2)=p^\mu p_{(\beta}\Pi_{ \alpha) \mu}(p^2) =0           \end{align}
since $\Pi_{\mu \alpha }(p^2)=-p^2\theta_{\mu \alpha}\tilde \pi(p^2)$ is the polarization tensor for the vector QED$_4$. This result indicates that the radiative corrections are proportional to the ${P^{(1)}_{ss} }_{\ \ \rho \sigma}^{\mu \nu}$ spin projector. This fact has important implications for the renormalizability of the theory.\\
\indent The self-energy for the fermion field is the following
\bea 
     i\Sigma(p)=-e^2\int \frac{d^4k}{(2\pi)^4}\frac{(p-k)^{(\mu}\gamma^{\nu)}i(\gamma^\beta k_\beta+m_e)(i)\mathcal{G}_{\mu \nu \alpha \beta}\big[(p-k) \big ](p-k)^{(\alpha}\gamma^{\beta)}}{\big( k^2-m^2_e\big)} \eea

%SELF ENERGY quando tomar cutkosky gera 2Im self. e surge um menos devido a i ao quadrado.
\indent Working out this expression, one gets
\begin{align} i\Sigma(p)=-e^2\int \frac{d^4k}{(2\pi)^4}\frac{\gamma^\mu(\gamma^\beta k_\beta+m_e)}{\big( k^2-m^2_e\big)}\frac{ \big(\eta_{\mu \beta}-\omega_{\mu \beta}+\omega_{\mu \beta}/\tilde \lambda    \big)\gamma^\beta}{(p-k)^2}\end{align}
which recovers the same expression from the usual QED$_4$ in Feynman gauge, if the choice $\tilde \lambda=1$ is considered.\\
\indent All QED$_4$ amplitudes are indeed recovered by this model due to the external states prescription and the fact that the internal bosonic lines associated with all the radiative corrections contribute as 
\begin{eqnarray}    p^{(\mu}\gamma^{\nu)}\mathcal{G}_{\mu \nu \alpha \beta} (p)\mathcal{O}p^{(\alpha}\gamma^{\beta)} \end{eqnarray}
\indent The $p_\mu$ denotes the momentum flowing in this internal bosonic line and $\mathcal{O}$ has spinor indices and dependence on integration and external momentum, being related to the fermion propagator. Then, due to the Gordon identity, just the $P_{SS}^{(1)}$ sector of the boson propagator leads to physical contributions, reproducing the standard vector electrodynamics result.\\

\subsection{On the Renormalizability of the Tensor Electrodynamics}

\indent There is still a question on renormalizability. One cannot take it for granted, considering the complicated tensor structure of the bosonic two-point part. The structure of the polarization tensor as well as its derivative order are equally important in this respect. The renormalized one-loop bosonic two-point part of the quantum action reads \footnote{For a quantum field generically denoted as $\mathcal{Q}_A(x)$, we consider its Fourier transform as $\mathcal{Q}_A(x)=\int \frac{d^4p}{(2\pi)^4}\mathcal{Q}_A(p)e^{-ip.x}$.}
\begin{align} \Gamma^R[\mathcal{N}]=\frac{1}{2(2\pi)^4}\int d^4p\ \mathcal{N}^{\mu \nu\ R}(p)\Big\{&-\frac{1}{2}(p^2)^2\big(1+\delta_\mathcal{N}+\tilde \pi(p^2) \big) P^{(1)}_{ss} \nonumber \\&+\tilde \lambda^R p^4P^{(0)}_{\omega \omega}+\lambda^R \big(p^8 P^{(2)}_{ss}+p^8P^{(0)}_{ss}\big)   \Big\}_{\mu \nu \alpha \beta}\mathcal{N}^{\alpha \beta \ R}(-p) \nonumber \\    \label{e}     \end{align}

\noindent in which the following relations for renormalized fields $\mathcal{Q}_A(x)=\sqrt{Z_A}\mathcal{Q}_A^R(x)$ and for the parameters $\mathcal{M}_C(x)=Z_C \mathcal{M}_C^R(x) $ are considered. The
constraints between the renormalization constants appearing for the massless limit of the model studied in Sec. \ref{sec10} are considered here. We assume the definition $\delta_\mathcal{N}=Z_\mathcal{N}-1$ throughout.\\
\indent Using dimensional regularization, it is possible to derive the explicit one-loop contribution \cite{xxxsch}
\bea  \tilde \pi(p^2)=\lim_{\epsilon \to 0}\frac{e^2}{2\pi^2}\int_0^1 dx  \big(1-x\big)x\Big[ \frac{2}{\epsilon}+\ln\Big(\frac{4\pi e^{-\gamma_E}\mu^2}{m^2_e-p^2x(1-x)}   \Big) \Big] \label{pi}\eea
with $\mu$ representing an energy scale and $\gamma_E$ being the Euler-Mascheroni constant. $\epsilon\to 0$ is defined as  $D=4-\epsilon$.\\
\indent The renormalized saturated propagator reads 
\bea
     T^{*\mu \nu}(k)\mathcal{G}_{\mu \nu \alpha \beta}^R(k)T^{\alpha \beta}(k)=-T^{*\mu \nu}(k)\Bigg\{\frac{2P^{(1)}_{ss}}{k^2\big(k^2(1+\delta_\mathcal{N}+\tilde \pi(k^2))\big)}\Bigg\}_{\mu \nu \alpha \beta} T^{\alpha \beta}(k)       \eea
     with c-number external sources defined as $T_{\mu \nu}(k)=\frac{1}{2}\big(k_\mu J_\nu(k)+k_\nu J_\mu(k)\big)$, with $k_\mu J^\mu(k)=0$ in accordance with the symmetries of the model.\\
\indent The on-shell renormalization condition for the propagator consists in demanding a massless pole for the saturated amplitude with residue equal to $-|J_\mu(k)|^2$, even in the interacting case, being positive definite. This condition implies in $\delta_\mathcal{N}=-\tilde \pi(0)$, meaning that the physical requirements are enough to generate a renormalized action with no divergent pieces. The claim on the amplitude positivity can be easily verified considering the source's transverse nature and the appropriate massless momentum frame such as $k_\mu=(k,k,0,0)$. \\
\indent Now for the sake of comparison, consider the spin $1$ model of the second derivative order of eq. (\ref{model}) with the interaction Lagrangian below
\bea    \mathcal{L}_I=\frac{\sqrt{2}}{m}\p^\nu H_{\mu \nu}J^\nu+\frac{J_\mu J^\mu}{2m^2}\eea
\indent It is possible to prove that it is not entirely equivalent to the Proca electrodynamics if, in addition to the c-number sources terms due to the quantum variational principle, a non-linear field interaction in which a source with a spinor structure $\Bar \Psi(x)\gamma_\mu \Psi(x) $ is also included. This and the Proca model present profound differences associated with their renormalization properties.\\
\indent In order to clarify this last point, an auxiliary field can be introduced to write {fermion self-interaction part in a Hubbard-Stratonovich transformation}
{\bea \frac{\Bar \Psi(x)\gamma_\mu \Psi(x)  \Bar \Psi(x)\gamma^\mu \Psi(x) }{2m^2}=-\frac{m^2}{2}\Phi_\mu \Phi^\mu+\Phi_\mu \Bar \Psi(x)\gamma^\mu \Psi(x) \eea}
 up to Gaussian integration, being implicitly assumed that $\mathcal{D}\Phi_\mu$ is added in the associated path integral measure. \\
 \indent Integrating out the fermion fields and ignoring the external source terms for a while, the renormalized two-point part of the bosonic sector of the quantum action reads
\begin{align}     &\Gamma^R\big[H\big]=\frac{1}{2(2\pi)^4}\int d^4p \Bigg\{\big|p^\mu H_{\mu \nu}(p)\big|^2    +\frac{m^2}{2}\Big(\big|H_{\mu \nu}(p)\big|^2-\frac{\big|H(p)\big|^2}{3}          \Big)    -\frac{m^2}{2}\Phi_\mu(p) \Phi^\mu(-p)\nonumber \\&+\Phi_{\mu}(p)p^2\theta^{\mu \nu}\tilde \pi(p)\Phi_\nu(-p)+\frac{2p^2}{m^2}\tilde \pi \big|p^\mu H_{\mu \nu}(p)\big|^2+\frac{2\sqrt{2}}{m}\Phi_{\mu}(p)p^2\theta^{\mu \nu}\tilde \pi(p)p^\alpha H_{\alpha \nu}(-p) \Bigg\}\end{align}

\indent If one tries to renormalize this system, extra conditions, beyond the physical ones, are necessary to eliminate all divergent terms from the action even if the $\Phi_\mu(x)$ field were absent. Namely, counter terms different from the bare Lagrangian ones must be considered to achieve this goal, something that is generally associated with a non-renormalizable system. A deeper analysis of this last point is performed in the recent paper \cite{oscar}, with a narrower characterization of this discussion. The illustrative particular example of the scalar QED$_4$ shows that a naive approach to renormalizability can lead to errors. Long ago \cite{rohr}, it was demonstrated that even being power-counting renormalizable, ultra-violate finiteness implies the necessity of adding a quartic scalar self-interaction. One can also address this question by the casual approach, see \cite{scharf}.

It is interesting to mention that the bosonic internal lines for the complete fermion self-energy are the sum of the tensor propagator and  the auxiliary vector propagator contracted with their respective vertex structures
\bea
       \gamma^\mu\mathcal{G}_{\mu \nu}^{(0)\Phi}(p)\mathcal{O}\gamma^\nu+\Delta^{\mu \nu}_{\ \sigma \omega}p^\sigma \gamma^\omega \mathcal{O} \mathcal{P}^{H}_{\mu \nu \alpha \beta}(p)\Delta_{\tau \varepsilon}^{\alpha \beta}p^\tau \gamma^\varepsilon=\gamma^\mu \mathcal{O}P^{(0)phys.}_{\mu \nu}(p)\gamma^\nu \eea
with the first term associated with the auxiliary field propagator, the second one with the tensor field and the latter represents the physical part of the bare Proca model propagator. $\mathcal{O}$ has the same interpretation as in the previous section. We are not considering the longitudinal parts of the propagators since, due to the Gordon identity, they do not contribute to the physical amplitudes obtained by attaching the external states. \\ 
\indent It can be proved that their sum plays the role of the physical part of the Proca propagator, meaning that the complete fermion self-energy is the same as in the Proca theory. The operator $\Delta^{\mu \nu}_{\ \sigma \omega}=\frac{1}{2}\Big(\delta^\mu_\sigma \delta^\nu_\omega +\delta^\nu_\omega \delta^\mu_\sigma  \Big)$ denotes the symmetrized identity.\\
\indent Considering the external state prescription, one can prove that both the massless fourth and the massive second derivative order models are in accordance with the optical theorem at least in one loop order. However, this is just a necessary condition {for unitarity}. Then, the coefficient of the interaction term for this model implies that it keeps a unitary nature just for energies below the cutoff $\Lambda \approx \frac{m}{e}$, with $m$ being the mass of the spin $1$ mediator.\\
\indent Interestingly, although non-renormalizable when formulated in terms of the $\Phi_\mu(x)$ and $H_{\alpha \beta}(x)$ variables, the second derivative model is indeed dual to Proca model even considering the interacting phase. All $1$PI  functions have their external lines associated with the combination
\bea \Phi_\mu(x)+\frac{\sqrt{2}}{m} \p^\beta H_{\beta \mu}(x)        \eea
\indent Then, redefining variables 
\bea \Phi_\mu(x)\to \Phi_\mu(x)- \frac{\sqrt{2}}{m} \p^\beta H_{\beta \mu}(x)        \eea
and Gaussian integrating in $H_{\mu \nu}(x)$, one recovers the complete Proca quantum interacting action.\\

\subsection{Asymptotic modes for the non-renormalizable theory}

\indent A well-defined theory can also be established by ignoring the current squared term in the model with a second derivative structure for the bosonic sector. Since it defines a legitimate model by itself, one can abandon the duality discussion and then focus on the following structure for the coupling
\bea\frac{e}{M} \p^\beta H_{\beta \mu}(x)\Bar{\Psi}(x)\gamma^\mu \Psi(x) \eea
with $M$ not necessarily related to the boson mass $m$.\\
\indent Then, in order to renormalize the system, one should consider the following set of counterterms
\bea \Gamma_{C.T.}=  \int \frac{d^4p}{(2\pi)^4}\Bigg\{\ \frac{H^{\mu \nu\ }(p)}{4}\Bigg[   -{P^{(1)}_{SS}}\Big(p^2\delta Z_{H}-\frac{p^4}{M^2}(1+\tilde \pi(0))\Big)\Bigg]_{\mu \nu \alpha \beta}H^{\alpha \beta \ }(-p)\Bigg\}               \eea
                involving the divergent piece of the polarization tensor. Extra higher derivative terms are also suitably included in order to eliminate divergent structures and to keep the same asymptotic modes as in the free case, differing only by the presence of renormalized parameters. This feature is an indication of the non-renormalizable nature of the system.  Regarding the counterterms from $\Gamma_{C.T.}$, the label $R$ for the renormalized structures is omitted for simplicity of notation. \\
\indent Then, for the renormalized system $\Gamma^R[H]=\Gamma[H]+\Gamma_{C.T.}$, the momentum dependent poles are defined as
\bea             m^2_{\pm}= \frac{M^2}{2(1-\tilde{\pi}^{fin.}(p))}\Bigg[(1+\delta Z_H)\mp \sqrt{ (1+\delta Z_H)^2-\frac{4m^2}{M^2} [1-\tilde{\pi}^{fin}(p)]          }    \Bigg]                               \eea
 The superscript $\textit{fin.}$ denotes the finite function $\tilde{\pi}^{fin.}(p)=\tilde{\pi}(p)-\tilde{\pi}(0)$. \\
\indent Considering the perturbative nature of $\tilde{\pi}^{fin.}(p)$, $\delta Z_H$ and taking into account the hierarchy of scales\footnote{This parameter is chosen to be in this scale in order to enable the mechanism described in this subsection.} $M^2\gg m^2$, a useful approximation can be derived
\bea  m^2_+= m^2(1-\delta Z_H)\ , \ m^2_-(p)=M^2(1+\tilde{\pi}^{fin.}(p)) (1+\delta Z_H)             \eea
\indent The renormalized propagator can be displayed as 
\begin{align}
     &\mathcal{P}_{\mu \nu \alpha \lambda}^R(p)=\frac{M^2}{  (1-\tilde{\pi}^{fin.}(p)) }\Bigg[\frac{{P^{(1)}_{SS}}_{\ \mu\nu,\alpha\lambda}}{\big(p^2-m^2_+\big)\big(p^2-m^2_-(p)\big)}\Bigg]\nonumber \\ &+\frac{{P^{(2)}_{SS}}_{\ \mu\nu,\alpha\lambda}}{m^2}+\frac{{P^{(0)}_{SS}}_{\ \mu\nu,\alpha\lambda}}{m^2}\Big(-1+\frac{3p^2}{m^2}\Big)-\frac{\sqrt{3}}{m^2}\Bigg({P^{(2)}}_{WS}+{P^{(2)}}_{SW}\Bigg)_{\ \mu\nu,\alpha\lambda}\label{prop1}\end{align}

\indent Then, near the $m^2_+$ pole, the physical part of the propagator acquires the following form
\bea \mathcal{P}_{\mu \nu \alpha \lambda}^R(p)=\frac{M^2}{  [1-\tilde{\pi}^{fin.}(m^2_+)] }\Bigg[\frac{{P^{(1)}_{SS}}_{\ \mu\nu,\alpha\lambda}}{\big(p^2-m^2_+\big)\big(m^2_+-m^2_-(m^2_+)\big)}\Bigg]\label{prop2}\eea 

\indent The unit residue condition can be achieved by a suitable fixation of $\delta Z_H$. \\
\indent Accordingly, choosing a counterterm parameter $M^2$ such that $M^2>4m^2_e$, the self-energy $\pi^{fin.}(p^2)$ evaluated at the extra pole $\tilde{m}^2_p\approx M^2$ contains an imaginary part. This extra pole explicitly defined below
\bea \tilde{m}^2_p=M^2(1+\mathcal{R}[\tilde{\pi}^{fin.}(\tilde{m}^2_p)]  )(1+\delta Z_H)   \eea
is shifted as
\bea \tilde{m}^2_p+i\gamma                       \eea
with $\gamma=M^2\Im[\tilde{\pi}^{fin.}(\tilde{m}^2_p)](1+\delta Z_H)>0$. The  $\mathcal{R} $ and  $\Im$ operators take the real and imaginary parts of a function, respectively. \\
\indent Since the associated residue is negative and $\gamma>0$, this extra excitation becomes a Merlin mode \cite{merline}. This is a special kind of resonance characterized by a ghost-like residue and a specific sign for the imaginary part of the pole. This is a backward traveling mode. Being unstable, it does not contribute to the asymptotic spectrum, playing no role in the unitarity constraints of the theory. This is an important mechanism for non-renormalizable theories, including some gravity models \cite{merlin2,merlin3}. For these systems, the radiative corrections may induce the occurrence of higher derivative terms possibly implying extra poles with problematic negative norms. However, as demonstrated, for some classes of interacting theories,  these poles become irrelevant for the asymptotic renormalized structure.

  \section{On the Possibility of New Couplings} \label{sec4}

  \indent A practical advance of studying dual relations between field theories is the possibility of new extensions. Starting from the established dual relation with QED$_4$, the proper structure of one of these formulations suggests new possibilities due to both the tensor structure and the higher derivative order of the model.\\
  \indent Considering the present case, due to the tensor structure for the bosonic sector, new couplings are possible, in principle. Regarding the present search, the guiding criteria are defined by renormalizability and gauge symmetry requirements.\\
  \indent It is possible to add a local Weyl invariant quartic interaction term, compatible with minimal coupling prescription, for this Abelian model, something that is not possible for the case of a vector field

\bea  \mathcal{L}_I=\frac{\beta}{4!}\Big(\big( \p^\nu \mathcal{N}_{\nu \alpha}-\frac{1}{4}\p_\alpha \mathcal{N}\big)\big( \p_\beta \mathcal{N}^{\beta \alpha}-\frac{1}{4}\p^\alpha \mathcal{N}\big)\Big)^2\eea
The interaction vertex is the following 
\begin{multline}
3\mathcal{V}^{\vartheta \epsilon \sigma \chi}_{\mu \nu \Sigma \omega}(k,m,h,l)=\beta \Big\{\Big( \Delta_{\mu \nu}^{\alpha \beta}k_\alpha-k^\beta \frac{\eta_{\mu \nu}}{4}   \Big)\Big( \Delta_{\tau \beta}^{\sigma \chi}m^\tau-m_\beta \frac{\eta^{\sigma \chi}}{4}   \Big) \Big( \Delta_{\Sigma \omega}^{\alpha \gamma}h_\alpha-h^\gamma \frac{\eta_{\Sigma \omega}}{4}   \Big)\Big( \Delta^{\vartheta  \epsilon}_{\rho \gamma}l^\rho-l_\gamma \frac{\eta^{\vartheta \epsilon}}{4}   \Big)\\ +\Big( \Delta_{\mu \nu}^{\alpha \gamma}k_\alpha-k^\gamma \frac{\eta_{\mu \nu}}{4}   \Big)\Big( \Delta_{\tau \beta}^{\sigma \chi}m^\tau-m_\beta \frac{\eta^{\sigma \chi}}{4}   \Big) \Big( \Delta_{\Sigma \omega}^{\alpha \beta}h_\alpha-h^\beta \frac{\eta_{\Sigma \omega}}{4}   \Big)\Big( \Delta^{\vartheta  \epsilon}_{\rho \gamma}l^\rho-l_\gamma \frac{\eta^{\vartheta \epsilon}}{4}   \Big)\\ +\Big( \Delta_{\mu \nu \alpha \beta}k^\alpha-k_\beta \frac{\eta_{\mu \nu}}{4}   \Big)\Big( \Delta^{\tau \gamma\sigma \chi}m_\tau-m^\gamma \frac{\eta^{\sigma \chi}}{4}   \Big) \Big( \Delta_{\Sigma \omega}^{\alpha \beta}h_\alpha-h^\beta \frac{\eta_{\Sigma \omega}}{4}   \Big)\Big( \Delta^{\vartheta  \epsilon}_{\rho \gamma}l^\rho-l_\gamma \frac{\eta^{\vartheta \epsilon}}{4}   \Big)\Big\}\end{multline}
in which $m_\beta$, $k_\beta$, $l_\beta$, and $h_\beta$ denote the momentum flow in its corresponding line.\\
\indent Interestingly, this interaction seems to be renormalizable. The associated coefficient is dimensionless and the superficial degree of divergence of the diagrams composed by just this kind of vertex does not increase with its number. It is given by $D=4L-4P_\gamma+\big(4V-N_\gamma    \big)$. The last term is due to the presence of $4$ derivatives in each vertex distributed into the four boson lines. Using the topological relation $L=P_\gamma-V+1$, one gets $D=4-N_\gamma$.\\
\indent A problematic fact is the partial violation of the reducible symmetry. It may lead to a violation of the spin $1$ nature of the model due to the form of the radiative corrections. It is then associated with both unitarity violation and the rise of divergent terms that are different than the original ones present in the bare Lagrangian. Although it has some good renormalization properties, the interaction seems to be not fully renormalizable. \\
\indent {The local gauge symmetry is broken implying that the gauge fixing part of the free propagator may contribute to the physical amplitudes. One way to deal with that is to suitably constrain the path integral to avoid non-desired gauge redundant contributions to the physical outcomes, keeping the spin 1 content of the model.\\
\indent Then, considering a slightly modified gauge fixing Lagrangian \footnote{Namely, replacing $\tilde \lambda$ by $ \tilde \lambda k^4$ and fixing $\lambda =\tilde \lambda$}} as compared to eq. \eqref{citar}, to have an invertible two-point structure \footnote{The local symmetry breaking is just in the self-interaction {term}}, and supplementing it with an extra traceless condition imposed by a Lagrange multiplier $\gamma(x)$,  {one can define a deformed theory by the inclusion of the extra Lagrangian contribution}
\bea \mathcal{L}_{extra}= \mathcal{N}^{\mu \nu}T_{\mu \nu}+\bar \Psi \eta+\bar \eta \Psi +\gamma \eta^{\mu \nu} \mathcal{N}_{\mu \nu} \eea
leading to a well-defined functional generator after adding external sources for every field.\\ \indent Although the quantum action has indeed potentially problematic terms that violate part of the reducible symmetry, the connected Green function generator $\mathcal{W}(T_{\mu \nu},\bar \eta,\eta)$, defined as $Z=e^{i\mathcal{W}}$, displays the effect of the integration of the Lagrange multiplier fields implementing the constraint in all quantum paths. The {connected} Green functions arising from source variations such as the free propagator 
\footnote{We denote by $\mathcal{W}^{(0)}$ the connected Green function generator in the free limit {$\beta \to 0$}. $\tilde{\delta}^{\mu \nu \chi \rho}\equiv \big( \Delta^{\mu \nu \chi \rho}-\frac{1}{4}\eta^{\mu \nu}\eta^{\chi \rho}\big)$ denotes the traceless projector with $\Delta^{\mu \nu \chi \rho}$ being the symmetrized identity. {The propagator $\mathcal{G}_{\chi \rho \sigma \gamma}(k)$ is the same as the one presented in the previous section but with a specific relation between the gauge fixing parameters}. }

\bea \frac{\delta^2 \mathcal{W}^{(0)}}{\delta T_{\mu \nu}(k)\delta T_{\alpha \beta}(-k)}= \tilde{\delta}^{\mu \nu \chi \rho}\mathcal{G}_{\chi \rho \sigma \gamma}(k)\tilde{\delta}^{\sigma \gamma \alpha \beta}         \eea
are all projected in the traceless sector, {avoiding any possibility of coupling with non-physical modes. According to the chapter $14.3$ of \cite{xxxsch}, the form of the previous equation indeed implies this mentioned projection for all the remaining $n$-point functions. }
The physical sector is indeed the same as in the previous sections.  Then, one easily notes that the bosonic internal lines are such that the reducible symmetry-breaking term in the vertex has no contributions when attached to it. Regarding graphs with external bosonic lines, the only source of symmetry violation is represented by the vertices directly attached to them. However, despite these undesired contributions to the quantum action, considering the constraints from the auxiliary field sector, the associated connected Green functions for the full interacting theory, encoding all the physics, can be consistently renormalized with no symmetry-breaking patterns at all. In summary, although the higher-rank nature of the field may lead to wider possibilities of couplings, these may violate the local symmetries. In the present case, the symmetry associated with the minimal coupling prescription is preserved while the hidden one is broken. Then, in order to keep the number of degrees of freedom, a suitable constraint should be added to the path integral measure. The intrinsic properties of the tensor representation are such that this process can be consistently done in terms of just local field functionals.  \\
\indent Accordingly, as we are going to see, a possible mass term with such Weyl invariant structure can be obtained from a non-invertible transformation analogous to the one that relates the so-called TDIFF to the WTDIFF spin $2$ models \cite{tdiff1,tdiff2}. For our specific model, this replacement keeps the spin $1$ structure while yielding a gauge invariant mass term, leading to infrared improvements.

 \section{A Gauge Invariant Mass Term and Propagating Modes}\label{sec5}
 
\indent This section is devoted to providing a semi-classical analysis of the Weyl invariant massive phase. The associated action reads

\bea S_M=\frac{1}{2}\int d^4x\Big(\p^\alpha \mathcal{N}_{\mu \alpha}\big(\eta^{\mu \nu}\Box-\p^\mu \p^\nu       \big)\p^\alpha \mathcal{N}_{\nu \alpha}+M^2\big( \p^\nu \mathcal{N}_{\nu \alpha}-\frac{1}{4}\p_\alpha \mathcal{N}\big)^2\Big) \nonumber \\       \eea
As mentioned, it can be obtained from the massive tensor model displayed in \cite{xxxd1}  through a non-invertible transformation
\bea \mathcal{N}_{\mu \nu}(x) \to \mathcal{N}_{\mu \nu}(x)-\frac{\eta_{\mu \nu}}{4}\mathcal{N}(x) \eea
enlarging the set of local symmetries. It also indicates that the associated quantization must present intricate features regarding the gauge fixing part, the measure, and the definition of the correct set of ghosts. This transformation introduces a Weyl symmetry while breaking part of the reducible one. After all, it replaces one hidden scalar symmetry by another scalar one associated with minimal coupling prescription. \\ 
\indent Due to the inclusion of this new term, just a residual part of the reducible symmetry is preserved, the ones whose associated local parameter obeys
\bea  \varepsilon^{\gamma \sigma \beta \rho}\varepsilon_{\gamma \mu \nu \alpha}\p_\sigma \p^\mu \Lambda^{\nu \alpha}_{\ \ \beta \rho}=0 \eea
representing a set of transverse and traceless symmetry transformations.\\
\indent More specifically, the reducible symmetry sector generated by the $P_{SS}^{(0)}$ projector is broken, while the one associated with $P_{SS}^{(2)}$ is preserved. \\
\indent In this case, the residual set of symmetries implies the following structure for the c-number external sources \footnote{Which also furnish hints for the possible couplings with fermions. } \bea T_{\mu \nu}(x)=\eta_{\mu \nu}R(x)+\p_{(\mu}J_{\nu)}(x)\label{const1}\eea
with the current $J_\mu(x)$ being not necessarily conserved.\\
\indent The Weyl symmetry imposes
\bea   4R(x)+\p^\mu J_\mu(x)=0  \label{const}             \eea

\indent Considering a source term projected in the standard spin $1$  structure ($R(x)=0$), the variational principle leads to the following equations of motion
\begin{align}  -&\frac{1}{2}\p_\alpha \Big(\big(\Box+M^2\big)\p^\beta \mathcal{N}_{\nu \beta}(x)- \p_\nu \big(\p_\mu \p_\sigma \mathcal{N}^{\mu \sigma}(x)  \big)\Big)-\frac{1}{2}\p_\nu \Big(\big(\Box+M^2\big)\p^\beta \mathcal{N}_{\alpha \beta}(x)-\p_\alpha\big(\p_\mu \p_\sigma \mathcal{N}^{\mu \sigma}(x)  \big)\Big)\nonumber  \\
&-\frac{M^2\Box }{16}\mathcal{N}\eta_{\alpha\nu}+
\frac{1}{4}\Big(M^2\eta_{\alpha \nu}\p^\gamma \p^\chi \mathcal{N}_{\gamma \chi} +M^2\p_\alpha \p_\nu \mathcal{N}   \Big)+\p_{(\nu}J_{\alpha)T}(x)=0\nonumber \\  \label{g} \end{align}
with a source term $T_{\nu\alpha}(x)=\p_{(\nu}J_{\alpha)T}(x)$, in which $T$ denotes a transverse vector,  added to the system.\\
\indent The choice for the gauge condition for Weyl symmetry is $\p^\gamma \p^\chi \mathcal{N}_{\gamma \chi}(x)=0$, while for the reducible one, it reads
\bea \Box^2\Big(P_{SS}^{(2)}\Big)_{\mu \nu}^{\ \alpha \beta}\mathcal{N}_{\alpha \beta}(x)=0\eea
suitable to reduce the freedom associated with the transverse and traceless unphysical modes.\\
\indent Applying the $P^{(1)}_{ss}$ spin $1$ projector to this equation, the spin-$1$ part is separated
\bea    k_\alpha \Big(\big(-k^2+M^2\big)k^\beta \mathcal{N}_{\nu \beta}(k)\Big)+k_\nu \Big(\big(-k^2+M^2\big)k^\beta \mathcal{N}_{\alpha \beta}(k)\Big)=2ik_{(\nu}J_{\alpha)T}(k) \eea
in a procedure conveniently performed in momentum space.\\
\indent Contracting the gradient $\p^\alpha$ with the equation \eqref{g} leads to a harmonic trace
\bea   \Box \mathcal{N}(x)=0                          \eea

\indent Considering the residual symmetries for the Weyl and reducible transformations associated with harmonic parameters, both the trace and the transverse and traceless sectors can be eliminated. This possibility is strongly dependent on the condition $R(x)=0$.\\
\indent Therefore, considering again the momentum space, we define the following vector  $k^\beta \mathcal{N}_{\nu \beta}(k)\equiv V_\nu(k) $, with $k_\mu V^\mu(k)=0$. Taking into account the constraints, we get the spin $1$ structure for the tensor field. For the case of a free asymptotic excitation, one gets
\bea \mathcal{N}_{\nu \beta}(k)=\frac{k_\nu V_\beta(k)+k_\beta V_\nu(k)}{M^2} \eea

\indent Therefore, even with this partial symmetry-breaking mass term, the model keeps its spin-$1$ nature if this specific source structure with $R(x)=0$ is considered. The Lagrangian mass term implies that the non-observable scalar sector of the reducible symmetry is replaced by the Weyl symmetry associated with the minimal coupling prescription.\\
\indent The associated gauge-fixed Lagrangian can be conveniently written in terms of the spin projectors

\begin{align} S_M=\frac{1}{2}\int d^4x\mathcal{N}^{\mu \nu}\Big(&-\frac{1}{2}\Box(\Box+M^2) P^{(1)}_{ss} +\big(\tilde \lambda \Box^2 -\frac{9M^2\Box}{16}     \big) P^{(0)}_{\omega \omega}+\lambda \big(\Box^4P^{(2)}_{ss}\big)-\frac{3M^2\Box}{16}P^{(0)}_{ss}\nonumber \\&+\frac{3\sqrt{3} \Box M^2}{16} \big(P^{(0)}_{\omega s}+P^{(0)}_{s \omega }\big) \Big)_{\mu \nu \alpha \beta}\mathcal{N}^{\alpha \beta} \nonumber \\  \label{u}       \end{align}

\noindent implying  the following form for the propagator
\bea \mathcal{G}_{\mu \nu \alpha \beta}=\Big(\frac{1}{\lambda k^8} P^{(2)}_{ss}-\frac{2}{k^2(k^2-M^2)}P^{(1)}_{ss} + \frac{3(1+\frac{16k^2\tilde \lambda}{9})}{\tilde \lambda k^4}P^{(0)}_{ss}      +\frac{1}{\tilde \lambda k^4}P^{(0)}_{\omega \omega}-\frac{\sqrt{3}}{\tilde \lambda k^4}\big(P^{(0)}_{\omega s}+P^{(0)}_{s \omega }\big)\Big)_{\mu \nu \alpha \beta} \nonumber \\        \eea

\indent The saturated amplitude from which we derive the necessary condition for tree-level unitarity \cite{saturated}, with the external sources  restricted to the conventional spin one structure ($R(x)=0$, see \eqref{const}) reads
\bea  \mathcal{A}(k)=iT^{*\mu \nu}(k)\ \mathcal{G}_{\mu \nu \alpha \beta} T^{\alpha \beta}(k)\eea                              
For an external source structure projected on the spin $1$ sector, $T^{\mu \nu}(k)=k^{(\mu}J^{\nu)}_T$, it is possible to derive the correct unitary condition for the residue of its imaginary part.  If $R(x)\neq 0$, the external sources are defined in \eqref{const1} and \eqref{const} leading to an amplitude that is also independent of the gauge fixing parameters but presents a problematic indefinite sign for the residue.

\subsection{On the observable propagating mode}

\indent Denoting the Weyl symmetry generator as ${\mathcal{Q}}^{Weyl}$, one gets
\bea  \Big[\mathcal{Q}^{Weyl}, \partial^\mu \mathcal{N}_{\mu \nu}(x)\Big]=\p_\nu\Lambda(x)       \eea
with $\Lambda(x)$ being a c-number field.\\
\indent Then, unless $\Lambda(x)$ is a constant, the symmetry is spontaneously broken and the system presents a discrete massless pole according to the Goldstone theorem \cite{gol62}. This pole is associated with the trace contribution.\\
\indent Therefore, the observable massive modes are defined as
\bea  \partial^\mu \mathcal{N}_{\mu \nu}^{phys.}\equiv  \partial^\mu \mathcal{N}_{\mu \nu}-\frac{\p_\nu \mathcal{N}}{4}    \eea
which reduces to the previously obtained physical field after considering the gauge fixing constraints.

\section{ Helicity Decomposition and Fundamental Classical  Aspects}\label{sec7}

\indent The helicity decomposition is a suitable Lagrangian method to investigate the propagating physical modes characterizing a given field theory model. It was originally formulated in the article \cite{heli1} with recent applications in a variety of contexts in physics \cite{heli2,heli3}.\\
\indent The fields  are decomposed in such a way that no extra time derivatives are added to keep the original canonical structure \footnote{{$\omega_{ij}$ is the spatial part of the longitudinal projector defined in the appendix A.}}
\begin{align}
    \mathcal{N}_{00}(x)=A(x), \quad \mathcal{N}_{0i}(x)=\p_iB(x)+v_i^T(x), \quad \mathcal{N}_{ij}(x)=\psi(x)\delta_{ij}+\omega_{ij}E(x)+h_{ij}^{TT}(x)+2\p_{(i}F_{j)}^T
\end{align}
\indent The free Lagrangian can be written in terms of these modes
\begin{multline}
    \mathcal{L}=\frac{1}{2}\Bigg\{  -\dot \phi \nabla^2\dot \phi+M^2\phi \nabla^2\phi -\chi \nabla^2\chi+M^2 \chi \chi-2(\nabla^2\phi)\dot \chi+\big(\dot v_i^T+\nabla^2F_i^T\big)\Big(\Box+M^2\Big)\big(\dot v_i^T+\nabla^2F_i^T\big)\Bigg\}   \label{helic}\end{multline}
with the Bardeen variables defined as $\chi(x)=\dot A(x)-\nabla^2B(x)$ and $\phi(x)=\dot B(x)-\big(  \psi(x)+E(x) \big) $.\\
\indent The trace $\mathcal{N}(x)$, a pure gauge field, was already eliminated by an appropriate choice of the Weyl transformation parameter implying in $A(x)=3\psi(x)+E(x)$, after a complete fixation of the symmetry. This mentioned transformation  is expressed in terms of the new variables as
\bea          \delta \psi(x)=-\Omega(x),\quad \delta A(x)=\Omega(x)      \eea
with $\Omega(x)$ being an arbitrary scalar field parameter associated with the Weyl symmetry which was fixated to yield \eqref{helic}. The residual invariance is then defined as the subgroup of the reducible gauge transformations preserving the trace.  \\
\indent The variations due to the reducible symmetry transformations are the following \footnote{TT denotes a transverse and traceless field.}
\begin{align}  &\delta A(x) =  \nabla^2\gamma(x),\quad \delta B(x)= \dot \gamma(x),\quad \delta \psi(x)=-\ddot \gamma(x), \quad  \delta E(x)=2\ddot \gamma(x)  \nonumber\\ & \delta F_i^T(x)=\dot \beta_i^T(x),\quad \delta v_i^T(x)=-\nabla^2\beta_i^T(x), \quad \delta h_{ij}^{TT}(x)=\mathcal{K}_{ij}^{TT}(x)             \end{align}
These are associated with the fact that the action is mainly written in terms of $\p^\mu \mathcal{N}_{\mu \nu}(x)$. The residual phase preserving the trace is defined below
\bea \Box \gamma(x)=0\eea

\indent Considering the information from the field equations of motion, the action acquires the form 
\begin{align}
    \mathcal{L}=-\frac{1}{2}\Bigg\{ \phi_n\Big(\Box+M^2\Big)\phi_n+        \mathcal{V}^T_i\Big(\Box+M^2\Big)\mathcal{V}^T_i \Bigg\}   \end{align}
with\footnote{The $\nabla^2$ operator has negative eigenvalues.} $\phi_n(x)=\sqrt{\Big(\frac{-\nabla^2}{-\nabla^2+M^2}\Big)}M\phi(x)$ and $\mathcal{V}^T_i(x)\equiv \dot v_i^T(x)$. We considered the fact that $F_i(x)$ and $h_{ij}^{TT}(x)$ are pure gauge variables. This analysis reveals that the system has indeed $3$ degrees of freedom, characteristic of a massive spin $1$ particle.

\section{Constraints from the Quantum Equations of Motion }\label{sec8}

\indent The previous analysis was a semi-classical one, leading to important suggestions. Taking this information into account, we investigate if these achievements are in accordance with the full quantum content of the theory.
The functional generator encoding this information reads \footnote{We are considering the notation associated to the product $\mathcal{D}\Phi_A\equiv \Pi_{A} \mathcal{D}\Phi_A(x)$ with $A$ denoting the $N$ independent components of a given field generically denoted as $\Phi_A(x)$.}
\begin{align}  Z\big[\eta,\bar \eta,T\big ]=\int &\mathcal{D}\mathcal{N}_{\mu \nu} \mathcal{D}\Psi \mathcal{D}\bar \Psi \exp i\int d^4x\Big[+i\bar \Psi\gamma^\mu \p_\mu \Psi-m_e\bar \Psi \Psi-\p^\nu \mathcal{N}_{\nu \mu}\big( e\bar \Psi\gamma^\mu\Psi+{\tilde e}\p^\mu (\bar \Psi \Psi) \big)  \nonumber \\&+{\tilde e}\mathcal{N}^{\mu \nu}\eta_{\mu \nu} \frac{\Box}{4}(\bar \Psi \Psi)+\frac{1}{2}\p^\alpha \mathcal{N}_{\mu \alpha}\big(\eta^{\mu \nu}\Box-\p^\mu \p^\nu       \big)\p^\alpha \mathcal{N}_{\nu \alpha}+\frac{M^2}{2}\Big( \p^\nu \mathcal{N}_{\nu \alpha}-\frac{1}{4}\p_\alpha \mathcal{N} \Big)^2 \nonumber \\&+T_{\nu \alpha}\mathcal{N}^{\nu \alpha}+\bar \eta \Psi+\bar \Psi \eta+\frac{\tilde \lambda}{2}(\p^\nu \p^\alpha \mathcal{N}_{\nu \alpha})^2+\lambda \mathcal{N}_{\mu \nu}\Box^4\Big[P^{(2)}_{SS}\Big]^{\mu \nu \alpha \beta}\mathcal{N}_{\alpha \beta}\Big]         \end{align}
with the most general second derivative order trilinear discrete symmetry-preserving spinor interaction allowed by the reduced set of symmetries. {Here, the coupling $\tilde e$ has the correct mass dimension to ensure a dimensionless action in natural units}. Namely, for a field-valued $R(x)\neq 0$. The c-number sources $\eta$, $\bar \eta$ and $T_{\mu \nu}$ are also added to provide the quantum variational principle.\\
\indent The quantum action can be obtained in terms of the connected Green function generator defined as $Z=e^{i\mathcal{W}}$. Then, after performing the Legendre transform \cite{xxxsch} 
\bea  \Gamma=\mathcal{W}-\int d^4x\Big(\bar \eta \Psi+\bar \Psi \eta+T_{\mu \nu}\mathcal{N}^{\mu \nu}   \Big) \eea adding the previously defined c-number source terms,  the quantum action is established.\\
\indent  From variations on the quantum equations of motion in the presence of a spinor interaction  whose respective $R(x)$ is given by  $R(x)={\tilde{e}} \frac{\Box}{4}\big(\bar \Psi(x) \Psi(x)\big)$ depicted above, it is possible to  derive the result bellow by the proper application of spin projectors
\begin{multline}   \theta^{\mu \nu}\frac{\delta^2\Gamma}{\delta \mathcal{N}_{\mu \nu}(x)\delta \mathcal{N}_{\chi \varepsilon}(y)}
+\frac{3e}{4}\Box^x  \int d^4ud^4\omega {\mathcal{S}}_{ac}(u-x) \frac{\delta^3\Gamma}{\delta \mathcal{N}_{\chi \epsilon}(y)\delta \bar \Psi_b(\omega) \delta \Psi_c(u) } {\mathcal{S}}_{ba}(u-x)  + .\ .\ .       =0 \end{multline}
 Here, $\theta_{\mu \nu}$ is the transverse projector defined in the appendix A. The terms in ellipsis denote gauge fixing and massive breaking terms from the quadratic part and $\mathcal{S}_{ab}(x,y)$ denotes the full fermion propagator with explicit spinor indices.\\
\indent It indicates that the bosonic inverse propagator receives radiative corrections implying that the renormalized quantum equation for the trace does not have a harmonic nature and cannot be eliminated from the physical spectrum by a proper gauge fixation. It leads to a new propagating mode, implying unitarity violation. On the other hand, if the operator $R(x)$ vanishes, the trace is indeed pure gauge even taking into account the full radiative corrections, in accordance with the semi-classical analysis. Therefore, in this case, the pole of the $P_{SS}^{(0)}$ sector keeps its non-observable nature even considering the full quantum corrections.\\
\indent The complete vertex function appearing above can be related to the one associated with the charge renormalization in momentum space as
\bea        \frac{\delta^3\Gamma}{\delta \mathcal{N}^{\chi \epsilon}(k)\delta \bar \Psi_b(\tilde p) \delta \Psi_c(p) }= \frac{\delta^3\Gamma}{\delta( k_\mu \mathcal{N}^{\mu \beta}(k))\delta \bar \Psi_b(\tilde p) \delta \Psi_c(p) }\Delta_{\chi \epsilon}^{\beta \sigma}k_\sigma\eea
according to the developments of Sec. \ref{sec10}.\\

\section{On the Quantum Stueckelberg Procedure for Higher Derivative Theory}\label{sec9}

\indent This section is devoted to proving that the Stueckelberg procedure \cite{xxxstu,stu2} can be successfully performed even for a higher derivative theory if one takes into account the whole quantum structure. Here, the focus is just on the two-point part of the bosonic Lagrangian. All the manipulations performed here do not influence the interaction term $\p^\mu \mathcal{N}_{\mu \nu}\bar \Psi(x)\gamma^\nu \Psi(x)$. Therefore, we highlight just the bosonic two-point part of the action.  We consider the Stueckelberg procedure for the scalar part of the reducible local symmetry broken by the mass term. We show that the violation of this symmetry is indeed responsible for the third extra polarization that typically arises for spin $1$ massive mediators. Therefore, one can choose to break this hidden symmetry and keep the one associated with the $U(1)$ minimal coupling prescription.\\
\indent The functional generator is related to the following action
\begin{align} S=&\frac{1}{2}\int d^4x\Big\{\p^\alpha \mathcal{N}_{\mu \alpha}\big(\eta^{\mu \nu}\Box-\p^\mu \p^\nu       \big)\p^\alpha \mathcal{N}_{\nu \alpha}+M^2\Big( \p^\nu \mathcal{N}_{\nu \alpha}-\frac{1}{4}\p_\alpha (\mathcal{N}+\frac{\varphi}{M})\Big)^2 \nonumber \\ &+\lambda'\mathcal{N}_{\mu \nu}\Box^4\Big[P^{(0)}_{SS}\Big]^{\mu \nu \alpha \beta}\mathcal{N}_{\alpha \beta}+\lambda \mathcal{N}_{\mu \nu}\Box^4\Big[P^{(2)}_{SS}\Big]^{\mu \nu \alpha \beta}\mathcal{N}_{\alpha \beta}+\tilde \lambda \big(\p^\mu \p^\nu \mathcal{N}_{\mu \nu}\big)^2+\bar C\ \Box^2 C  \Big\}    \end{align}
being implicitly assumed that $\mathcal{D}\varphi(x)$, associated with the Stueckelberg compensating field, as well as a ghost sector, is included in the measure.\\
\indent Since the theory is Abelian, just the ghosts associated with the $P_{SS}^{(0)}$ broken sector of the reducible local transformations are explicitly displayed here to highlight the full quantum content of the procedure. The gauge condition is associated with the $\lambda'$ parameter related to the fixation of the scalar reducible symmetry ensured by the compensating field.\\
\indent The presence of the scalar Stueckelberg compensating field guarantees the extra local freedom defined below
\bea \delta \varphi(x)=-M\sigma(x)\quad  ,\quad \delta \mathcal{N}_{\mu \nu}(x)= \frac{\theta_{\mu \nu}}{3}\sigma(x) \label{sym} \eea
associated with the mentioned scalar part of the reducible symmetry.\\
\indent Considering the limit $M\to 0$, the field $\varphi(x)$ is not pure gauge anymore. Then, one recovers  
\begin{align} S=&\frac{1}{2}\int d^4x\Big\{\p^\alpha \mathcal{N}_{\mu \alpha}\big(\eta^{\mu \nu}\Box-\p^\mu \p^\nu       \big)\p^\alpha \mathcal{N}_{\nu \alpha}+\frac{1}{16}\p_\mu \varphi \p^\mu \varphi +\lambda'\mathcal{N}_{\mu \nu}\Box^4\Big[P^{(0)}_{SS}\Big]^{\mu \nu \alpha \beta}\mathcal{N}_{\alpha \beta}\nonumber \\ &+\lambda \mathcal{N}_{\mu \nu}\Box^4\Big[P^{(2)}_{SS}\Big]^{\mu \nu \alpha \beta}\mathcal{N}_{\alpha \beta}+\tilde \lambda \big(\p^\mu \p^\nu \mathcal{N}_{\mu \nu}\big)^2+\bar C\ \Box^2 C  \Big\}    \end{align}
the massless action with just the ghosts related to the scalar reducible symmetry being explicitly displayed plus a unitary decoupled free scalar field Lagrangian. It means that the degrees of freedom are kept at the massless limit, and the third longitudinal polarization of the massive field is replaced by the free scalar field. Therefore, in order to include mass for a spin $1$ field, a scalar symmetry must be broken. Due to this specific tensor structure, the Weyl symmetry associated with a covariant derivative prescription can be maintained. Alternatively, one can also choose to break the Weyl symmetry and keep the completely reducible one, including its scalar part.   \\
\indent Now, we prove that the action with and without the Stueckelberg field yields the same result. Just two field redefinitions with unit Jacobian are necessary. The first one reads
\bea    \mathcal{N}_{\mu \nu}(x)\to \mathcal{N}_{\mu \nu}(x)-\frac{\theta_{\mu \nu}}{3}\frac{\varphi(x)}{M}                \eea
followed by
\bea        \varphi(x) \to \varphi(x) +M\theta^{\alpha \beta}\mathcal{N}_{\alpha \beta}(x)       \eea
\indent The remaining action terms associated with the compensating field becomes
\begin{align} S=&\frac{1}{2}\int d^4x\Big\{\frac{\lambda'}{M^2}\varphi(x)\Box^4\varphi(x)+\bar C\ \Box^2 C  \Big\}    \end{align}
\indent Considering the bosonic/Grassmann nature of the fields, a Gaussian integration leads to a constant normalization factor with no physical content. The relevance of this analysis is to show that no differential operator normalization is left by this procedure, meaning that it has no implications in the thermodynamics of the theory. The physical content of the theory is not altered by the Stueckelberg procedure even in the present non-standard situation. \\

\section{Ward-Takahashi Identities for the Massive Theory with Stueckelberg Fields}\label{sec10}
\indent The addition of the auxiliary field, implying a wider local freedom, is a useful procedure to derive constraints for the renormalization process via Ward-Takahashi-like identities.  The functional generator explicitly reads

\begin{align}  Z\big[\eta,\bar \eta,T_{\mu \nu},J\big ]=\int &\mathcal{D}\mathcal{N}_{\mu \nu} \mathcal{D}\varphi \mathcal{D}\Psi \mathcal{D}\bar \Psi \exp i\int d^4x\Big[+i\bar \Psi\gamma^\mu \p_\mu \Psi-m_e\bar \Psi \Psi-e\p^\nu N_{\nu \mu} \bar \Psi\gamma^\mu\Psi  \nonumber \\&+\frac{1}{2}\p^\alpha \mathcal{N}_{\mu \alpha}\big(\eta^{\mu \nu}\Box-\p^\mu \p^\nu       \big)\p^\alpha \mathcal{N}_{\nu \alpha}+\frac{M^2}{2}\Big( \p^\nu \mathcal{N}_{\nu \alpha}-\frac{1}{4}\p_\alpha (\mathcal{N}+\frac{\varphi}{M} )\Big)^2 \nonumber \\&+\varphi J+T_{\nu \alpha}\mathcal{N}^{\nu \alpha}+\bar \eta \Psi+\bar \Psi \eta+\frac{\tilde \lambda}{2}(\p^\nu \p^\alpha \mathcal{N}_{\nu \alpha})^2\nonumber \\&+\lambda'\mathcal{N}_{\mu \nu}\Box^4\Big[P^{(0)}_{SS}\Big]^{\mu \nu \alpha \beta}\mathcal{N}_{\alpha \beta}+\lambda \mathcal{N}_{\mu \nu}\Box^4\Big[P^{(2)}_{SS}\Big]^{\mu \nu \alpha \beta}\mathcal{N}_{\alpha \beta}\Big]         \end{align}
with the addition of c-number external sources for every quantum field, in order to derive a functional generator.\\
\indent The system is invariant under the following sets of  transformations,
\begin{align} \Psi(x)\to \Psi(x)+i\alpha(x)\Psi(x)\quad &,\quad  \bar \Psi(x)\to \bar \Psi(x)-i\alpha(x)\bar \Psi(x) \nonumber \\
\mathcal{N}_{\mu \nu }(x)\to \mathcal{N}_{\mu \nu }(x)+\frac{1}{e}\eta_{\mu \nu }\alpha(x)\quad& ,\quad \mathcal{N}^{\mu \nu}(x) \to \mathcal{N}^{\mu \nu}(x)+ \Big(P_{SS}^{(2)}\Big)^{\mu \nu}_{\ \alpha \beta}\mathcal{E}^{\alpha \beta}-\frac{\theta^{\mu \nu}}{3M}\Lambda(x)
\end{align}
in which $\alpha(x)$ is the parameter due to Weyl symmetry transformations. The functions $\mathcal{E}_{\mu \nu}(x)$ and $\Lambda(x)$ are the parameters associated with residual reducible symmetry and the broken one (recovered by the Stueckelberg procedure), respectively. The scalar field transforms as in equation \eqref{sym}, after a suitable change of notation for the symmetry parameter.\\
\indent From the invariance of the functional generator under the transformation generated by $P_{SS}^{(0)}$, it is possible to derive the following Ward-Takahashi identities (WTI)
\bea
    \frac{\theta_{\mu \nu}}{3}\frac{\delta^2\Gamma}{\delta \mathcal{N}_{\mu \nu}(k) \delta \mathcal{N}_{\beta \alpha}(-k) }+\lambda'\frac{\theta^{\beta \alpha}(k)}{3}-M\frac{\delta^2\Gamma}{\delta \varphi(k)\delta \mathcal{N}_{\beta \alpha}(-k)}=0
\eea
after passing to the momentum space with $\Gamma$ representing the effective action.
Then, considering the quantum equations for the compensating fields

\bea \frac{\delta^2\Gamma}{\delta \varphi(k)\delta \varphi(-k)}=\frac{k^2}{16} \quad ,\quad \frac{\delta^2\Gamma}{\delta \varphi(k)\delta \mathcal{N}_{\mu \nu}(-k)}=\frac{M}{4}\Big[\frac{k^2}{4}\eta_{\mu \nu}-k_\mu k_\nu \Big] \eea
which can be used to conclude
\bea
    \frac{\theta_{\mu \nu}}{3}\frac{\delta^2\Gamma}{\delta \mathcal{N}_{\mu \nu}(k) \delta \mathcal{N}_{\beta \alpha}(-k) }+\lambda'\frac{\theta^{\beta \alpha}(k)}{3}-\frac{M^2}{4}\Big[\frac{k^2}{4}\eta^{\alpha \beta}-k^\alpha k^\beta \Big] =0
\eea
a relation for some renormalization constants can be found $Z_{\mathcal{N}}Z_{\lambda'}=1=Z_{\mathcal{N}}Z_{M^2}=Z_\varphi$.\\
\indent Since the functional generator has Weyl symmetry, the following associated Ward-Takahashi identity can be found

\begin{align} \mathcal{S}^{-1}(\tilde p+k)-\mathcal{S}^{-1}(\tilde p)=\frac{k_\alpha}{e} \Gamma^\alpha(p,\tilde p,k=p-\tilde p) \nonumber \\       \end{align}
after passing to the momentum space
with $\mathcal{S}(k)$ denoting the complete fermion propagator and the vertex defined as 
\bea \Gamma^\alpha(p,\tilde p,k=p-\tilde p)\equiv \frac{\delta^3\Gamma}{\delta \bar \Psi(p)\delta \Psi(\tilde p)\delta (k^\beta \mathcal{N}_{\beta \alpha}(k))}  \eea
being the same as the one displayed in Sec. \ref{sec8}.\\
\indent Despite the alternative tensor structure, it leads to the same relation between charge renormalization and the bosonic wave function
\bea 1=Z_e\sqrt{Z_\mathcal{N}}\eea
as in QED$_4$ case. Then, the running of the charge is ruled just by the bosonic mediator renormalization. This is the reason why a proton and an electron have exactly the opposite charge, despite the fact that they participate in totally different kinds of interactions \cite{xxxsch}.\\
\indent The Weyl symmetry is also associated with the following identity for the propagator
\bea -\eta^{\mu \nu}\frac{\delta^2\Gamma}{\delta \mathcal{N}_{\mu \nu}(k) \delta \mathcal{N}_{\beta \alpha}(-k) }+\tilde \lambda k^2k_\beta k_\alpha=0 \eea
leading to $Z_\mathcal{N}Z_{\tilde \lambda}=1$.\\
\indent There is also the identity related to symmetry transformations generated by $P_{SS}^{(2)}$
\bea \Big[P^{(2)}_{SS}\Big]^{\mu \nu}_{\ \sigma \omega}\Bigg[ \frac{\delta^2\Gamma}{\delta \mathcal{N}^{\mu \nu}(k) \delta \mathcal{N}_{\beta \alpha}(-k) }+ \lambda (k^2)^4 \Delta_{\mu \nu }^{\alpha \beta}\Bigg]=0 \eea
leading to $Z_{\mathcal{N}}Z_\lambda=1$.\\

\section{Renormalization Conditions for the Massive Boson}\label{sec11}
\indent Considering the (WTI), the renormalized propagator becomes
\begin{multline}
     \mathcal{G}_{\mu \nu \alpha \beta}^R=\Bigg\{\frac{P^{(2)}_{ss}}{\lambda_R k^8} -\frac{2P^{(1)}_{ss}}{k^2\big(k^2(1+\delta_\mathcal{N}+\tilde \pi(k^2))-M^2_R\big)} + \frac{3(1+\frac{16k^2\tilde \lambda}{9})P^{(0)}_{ss}}{\tilde \lambda_R k^4}      +\frac{P^{(0)}_{\omega \omega}}{\tilde \lambda_R k^4}-\frac{\sqrt{3}\big(P^{(0)}_{\omega s}+P^{(0)}_{s \omega }\big)}{\tilde \lambda_R k^4}\Bigg\}_{\mu \nu \alpha \beta}        \end{multline}
The radiative corrections contribute by shifting the pole associated with the spin $1$ sector. Near the mass pole, the saturated propagator becomes
\\

\bea
    T^{*\mu \nu}(k) \mathcal{G}_{\mu \nu \alpha \beta}^{R\ (phys.)}(k)T^{\alpha \beta}(k)=- \frac{2T^{*\mu \nu}(k)(P^{(1)}_{ss})_{\mu \nu \alpha \beta}T^{\alpha \beta}(k)\Big(1-\frac{d(k^2\tilde{\pi}^R(k^2))}{dk^2}|_{m_p^2} \Big) }{k^2\big(k^2-m^2_p+im^2_P\Im \ 
\tilde \pi^R(m^2_p) \big)}         \eea
\bea        m_p^2=M^2_R-m^2_p \mathcal{R}\big(\tilde \pi(m^2_p)  \big)-m^2_p\delta_\mathcal{N}     \eea
Then, considering the explicit expression of \eqref{pi}, the renormalized radiative correction can be defined

\bea        \tilde{\pi}^R(k^2)= \tilde \pi(k^2) +\delta_\mathcal{N}     \eea
with the c-number external source $T_{\mu \nu}(k)=\frac{1}{2}\Big(k_\mu J_\nu+k_\nu J_\mu\Big)$, in which $k_\mu J^\mu=0$.\\
\indent Then, it is possible to define $\delta_\mathcal{N}$ by the requirement of obtaining $-|J_\mu|^2$ as the residue
\bea          \frac{d(k^2\tilde{\pi}^R(k^2))}{dk^2}|_{m_p^2}=0 \eea

\indent Therefore, if one wants to define an infrared improved QED$_4$ with no low energy singularities in the fermionic self-energy and vertex function, the range $M_p\lesssim 10^{-15}$eV must be considered. This very stringent bound is in accordance with recent data on the cosmological propagation of fast radio bursts \cite{radiob}. Since it is much smaller than the electron mass $m_e\approx 0,51 $ MeV, one concludes that  $\Im \ 
\tilde \pi^R(M^2_p)=0$.
 
\section{Conclusions and Future Perspectives}\label{sec12}

\indent Throughout this work, we obtained a gauge invariant massive extension of the electrodynamics. The discussion on duality also included non-linear field interactions. We developed a higher derivative tensor electrodynamics capable of reproducing all standard QED$_4$ amplitudes if a prescription for the external states is considered. All the Feynman rules and renormalization properties were derived and implemented in the generalized tensor structure. The higher rank tensor field allowed the obtainment of new couplings and a gauge invariant mass term. The symmetries and propagating degrees of freedom were investigated by a careful Stueckelberg analysis and also the helicity decomposition procedure. The constraints from the full quantum equations of motion were also considered. \\
\indent Although a robust analysis was performed, there are still some open questions. The derivation of the complete ghosts field Lagrangian, something necessary to define the thermodynamical properties of the system, is an interesting formal goal associated with the intricate structure of the residual local symmetries. Another question relies on the delimitation of the dualization procedure. For example, verifying if the dual correlation is preserved in a curved background or even in the presence of new interactions is a complementary objective.\\
\indent As mentioned in the introduction, the present achievements can be understood as a laboratory for some future perspectives in research on gravity. A feasible possibility consists in the search for a dual linearized gravity described through a rank three tensor field $\chi_{\mu \nu \alpha}(x)$, in terms of which the standard metric perturbation can be obtained through index contraction. The reason for such an approach is to look for a four-dimensional theory with full linearized diffeomorphism invariance \footnote{Enjoying not just the transverse but complete linearized diffeomorphism symmetry. } even considering the introduction of a mass term. \\
\indent In our specific perspective, based on a higher derivative theory described by a unique higher-rank tensor field, the objective is to achieve full master action duality with the standard Einstein-Hilbert model (EHM) for the non-interacting massless limit. This step depends on the avoidance of contact terms violating the dual relation with (EHM) Green functions in the Gaussian case. It would possibly lead to a well-defined prescription for extensions with good UV behavior (due to the absence of contact divergences). Then, it can be a consistent point of departure to include new effects like interactions and/or mass terms. \\
\indent The development of a class of renormalizable new interactions is another possibility associated with this approach. This goal also implies the necessity of the most general index symmetry to develop our procedure. Working out this new interaction can be a way to collaborate with research relating theoretical quantum gravity with cosmological demands such as the investigation of cosmic acceleration in the massive gravity framework \cite{mcosm}, for example.
As in this specific paper, the higher rank tensor structure can be applied to replace the linearized diffeomorphism symmetry breaking due to the mass term by the violation of a hidden local freedom, in analogy to the discussion on the quantum Stueckelberg trick in Sec. \ref{sec9}. Since our approach consists of deriving the full quantum properties and investigating duality under this perspective, a judicious knowledge of the propagator structure for rank $3$ fields is a necessary step. The articles on the spin $3$ theories \cite{spin} furnish a useful methodology for this future enterprise. The quantum Stueckelberg trick, renormalization conditions for the case of a field-dependent energy-momentum tensor interaction, and complete quantum constraints from Ward-Takahashi identities are correlated possibilities that generalize the present achievements.

\acknowledgments
\indent G.B. de Gracia thanks the São Paulo Research Foundation -- FAPESP  Post Doctoral grant No. 2021/12126-5. B.M. Pimentel thanks CNPq for partial support.

\section{Appendix A}

\indent The spin $1$ projectors, from which the rank $2$ ones are constructed, are listed below
\bea   \theta_{\mu \nu}(k)=\eta_{\mu \nu}-\frac{k_\mu k_\nu}{k^2}\quad , \quad \omega_{\mu \nu}= \frac{k_\mu k_\nu}{k^2}      \eea
With the properties
\bea \theta_{\mu \nu}(k)\theta^{\mu \alpha}(k)=\theta^{\alpha}_\nu(k)\,\
 \omega_{\mu \nu}(k)\omega^{\mu \alpha}(k)=\omega^{\alpha}_\nu(k)\ ,\ \theta_{\mu \nu}(k)\omega^{\mu \alpha}(k)=0\ ,\ \theta_{\mu \nu}(k)+\omega_{\mu \nu}(k)=\eta_{\mu \nu}\nonumber \\\eea
  Then, the spin $2$ projectors are of the following

\begin{multline}
      {P^{(2)}_{ss} }_{\ \ \rho \sigma}^{\mu \nu}=\frac{1}{2}\Big(\theta^{\mu}_{\ \rho}\theta^{\nu}_{\ \sigma}+\theta^{\mu}_{\ \sigma}\theta^{\nu}_{\ \rho} \Big)-\frac{\theta^{\mu \nu}\theta_{\rho \sigma}}{(D-1)}\quad ,\quad{P^{(1)}_{ss} }_{\ \ \rho \sigma}^{\mu \nu}=\frac{1}{2}\Big(\theta^{\mu}_{\ \rho}\omega^{\nu}_{\ \sigma}+\theta^{\mu}_{\ \sigma}\omega^{\nu}_{\ \rho}+\theta^{\nu}_{\ \rho}\omega^{\mu}_{\ \sigma}+\theta^{\nu}_{\ \sigma}\omega^{\mu}_{\ \rho} \Big)       \end{multline}

\bea {P^{(0)}_{s\omega } }_{\ \ \rho \sigma}^{\mu \nu}=\frac{\theta^{\mu \nu}\omega_{\rho \sigma}}{\sqrt{(D-1)}} \quad ,\quad {P^{(0)}_{ss} }_{\ \ \rho \sigma}^{\mu \nu}=\frac{\theta^{\mu \nu}\theta_{\rho \sigma}}{(D-1)}            \eea

\bea  {P^{(0)}_{\omega \omega} }_{\ \ \rho \sigma}^{\mu \nu}=\omega^{\mu \nu}\omega_{\rho \sigma} \quad,\quad  {P^{(0)}_{\omega s} }_{\ \ \rho \sigma}^{\mu \nu}=\frac{\omega^{\mu \nu}\theta_{\rho \sigma}}{\sqrt{(D-1)}}                     \eea

\bea    {P^{(0)}_{s\omega } }_{\ \ \rho \sigma}^{\mu \nu}=\frac{\theta^{\mu \nu}\omega_{\rho \sigma}}{\sqrt{(D-1)}}                                \eea

The completeness relations are the following

\bea  {P^{(2)}_{ss} }_{\ \ \rho \sigma}^{\mu \nu}+{P^{(1)}_{ss} }_{\ \ \rho \sigma}^{\mu \nu}+{P^{(0)}_{ss} }_{\ \ \rho \sigma}^{\mu \nu}+{P^{(0)}_{\omega \omega} }_{\ \ \rho \sigma}^{\mu \nu}=\frac{1}{2}\big( \delta^\mu_\rho \delta^\nu_\sigma+\delta^\nu_\rho \delta^\mu_\sigma      \big)                            \eea
The projector algebra reads $P^{(s)}_{IL}P^{(\tilde s}_{JK}=\delta^{s\tilde s}P^{(s)}_{IK}$

\section{Appendix B}

\indent The gamma matrices obey the relations
\bea \Big\{\gamma^\mu,\gamma^\nu  \Big\}=2I_{4\times 4}\eta^{\mu \nu }\eea
\indent  For the traces, we have
\bea
tr(\gamma^\mu)=0=tr(\gamma^{\mu \ 1 }...\gamma^{\mu \ 2n+1 }) \ , \ tr(\gamma^\mu\gamma^\nu)=4\eta^{\mu \nu }\eea
\bea tr(\gamma^\mu \gamma^\nu \gamma^\rho \gamma^\sigma)=
4\Big(\eta^{\mu \nu}\eta^{\rho \sigma}
-\eta^{\mu \rho} \eta^{\nu \sigma}+\eta^{\mu \sigma}\eta^{\nu \rho}\Big)\eea

\end{document}